\documentclass{jps-cp}
\usepackage{txfonts} 
\usepackage{lineno}
\usepackage{subfigure}

\title{Measurement of longitudinal spin transfer of the $\Lambda(\bar{\Lambda})$ hyperon in polarized $\rm{p}$$+$$\rm{p}$ collisions at $\sqrt{s}=200$ GeV at RHIC-STAR }

\author{Yi Yu for the STAR Collaboration}
\inst{Institute of Frontier and Interdisciplinary Science \& Key Laboratory of Particle Physics and Particle Irradiation (MOE), Shandong University, Qingdao, Shandong, 266237, China}

\email{201812156@mail.sdu.edu.cn}
\abst{Since the first surprising results on the spin structure of the proton by the EMC experiment in the late 1980s, 
much progress has been made in understanding the origin of the proton spin. 
However, the sea quark contribution to the proton spin, for example, 
the helicity distributions of the strange quark (anti-quark), $s(\bar{s})$, is still not well constrained by experimental data.
Since the $s(\bar{s})$ is expected to carry a substantial fraction of the spin of the $\Lambda(\bar{\Lambda})$ hyperon, 
measurements of the longitudinal spin transfer, $D_{LL}$, of the $\Lambda(\bar{\Lambda}$) hyperon can thus shed light 
on the helicity distribution of the $s(\bar{s})$ and the longitudinally polarized fragmentation functions.
In these proceedings, we will present the status of the $\Lambda(\bar{\Lambda}$) $D_{LL}$ analysis using  data collected at RHIC-STAR in 2015, 
for the pseudo-rapidity $|\eta| < 1.2$ and transverse momenta up to $8.0$ GeV$/c$. 
The $D_{LL}$ as a function of the longitudinal momentum fraction of the $\Lambda(\bar{\Lambda})$ hyperon in the jet is also investigated.
This data set is about twice as large as the 2009 data used for the previously published $D_{LL}$ results.
}

\kword{proton helicity structure, RHIC-STAR, $\Lambda$ hyperon, spin transfer}

\begin{document}
\maketitle


\section{Introduction}
Understanding the spin structure of the proton is one of the most challenging and fundamental questions in QCD. 
In 1988, the European Muon Collaboration published the puzzling result 
showing that the quark and anti-quark spin only contribute little to the spin of the proton\cite{EMC}. 
This result then inspired global experimental and theoretical studies in the last 30 years. 
For the helicity structure of the proton, the contribution from valence quarks is well constrained\cite{helicity_pdf}. 
However, the sea quark contributions, especially for the strange quark (anti-quark), $s(\bar{s})$, still have large uncertainties. 
Since the spin of the $\Lambda(\bar{\Lambda})$ hyperon is expected to be carried mostly by its valence $s(\bar{s})$, several theoretical studies\cite{theory_work,theory_work2, theory_work3, theory_work4, theory_work5} suggest that the $D_{LL}$ 
can provide constraints on the helicity distribution of $s(\bar{s})$ 
and the polarized fragmentation functions of $\Lambda(\bar{\Lambda})$. In particular, Ref.\cite{theory_work3} shows that measuring $D_{LL}$ as a function of the jet momentum fraction ($z$) carried by the $\Lambda(\bar{\Lambda})$ hyperon can directly probe the polarized jet fragmentation functions of $\Lambda$ and $\bar{\Lambda}$.  

In polarized p+p collisions, the longitudinal spin transfer of $\Lambda$ is defined as:
\begin{equation}
    D_{LL} \equiv \frac{\sigma_{p^+p\rightarrow \Lambda^+X}-\sigma_{p^+p\rightarrow \Lambda^-X}}{\sigma_{p^+p\rightarrow \Lambda^+X}+\sigma_{p^+p\rightarrow \Lambda^-X}}=\frac{\Delta \sigma}{\sigma},
\end{equation}
where ``$+$" or ``$-$" denotes the helicity of the proton or $\Lambda(\bar{\Lambda})$. 
In the factorization framework, the polarized cross section $\Delta \sigma$ can be expressed as 
the convolution of the helicity distributions, partonic cross sections and polarized fragmentation functions.

The Relativistic Heavy Ion Collider (RHIC) is the world's first and only polarized p+p collider 
which is capable of colliding the polarized proton beams at center of mass energies equal to 200 GeV and 510 GeV, 
making it an ideal facility for probing the spin structure of the proton. Previously, STAR has published the $D_{LL}$ results\cite{DLL_2018} using the data taken in 2009 with an integrated luminosity of 19 $\mathrm{pb}^{-1}$. In 2015, STAR recorded a larger data sample corresponding to an integrated luminosity of 52 $\mathrm{pb}^{-1}$ with an average beam polarization of about 54\%. With this data set, we performed more precise measurements on the $D_{LL}$ vs hyperon $p_T$ and the first measurements on the $D_{LL}$ vs hyperon momentum fraction ($z$) within jets.

\begin{figure}
\centering
\subfigure[$\Lambda$]
{\includegraphics[width=0.45\linewidth]{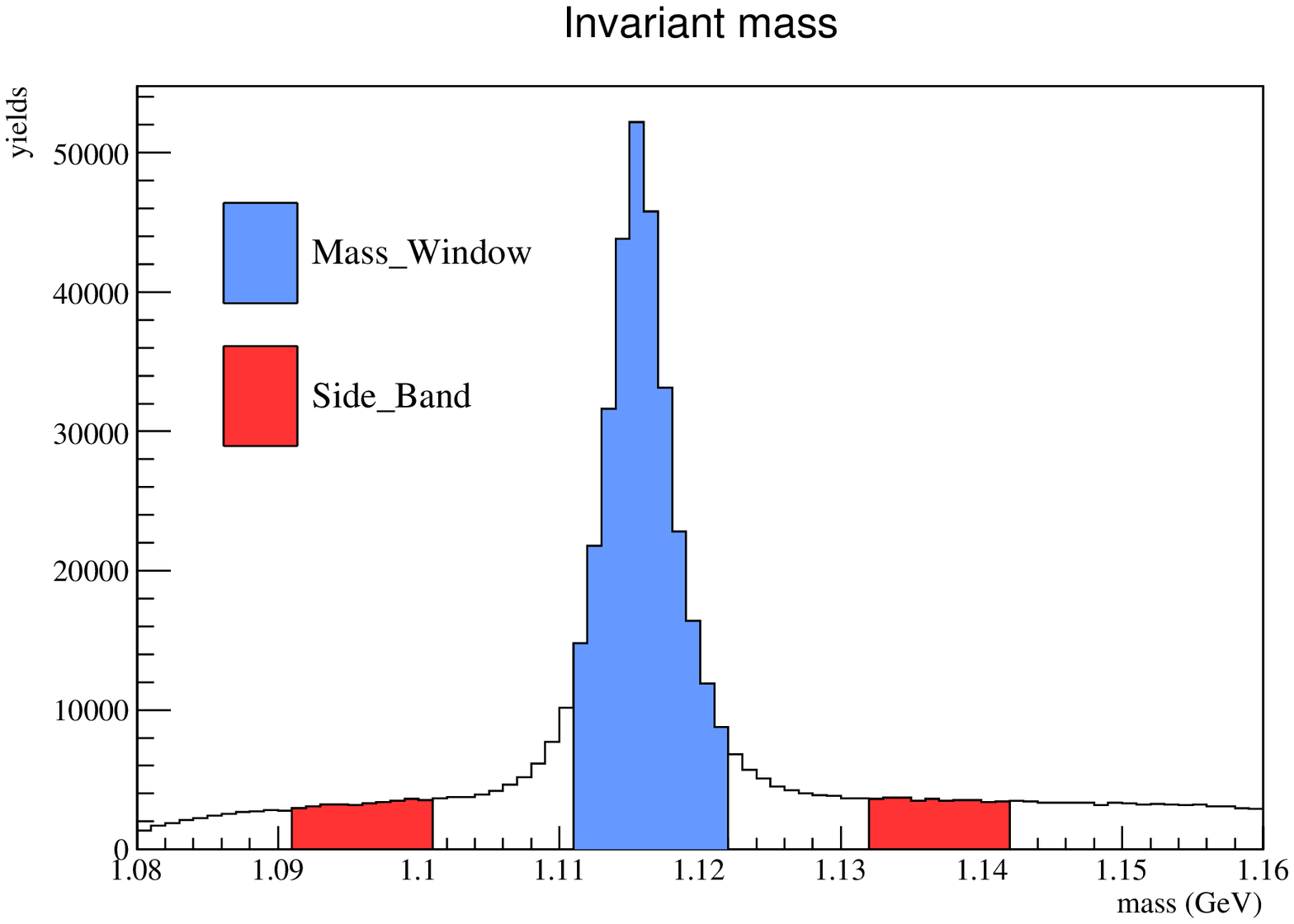}}
\centering
\subfigure[$\bar{\Lambda}$]
{\includegraphics[width=0.45\linewidth]{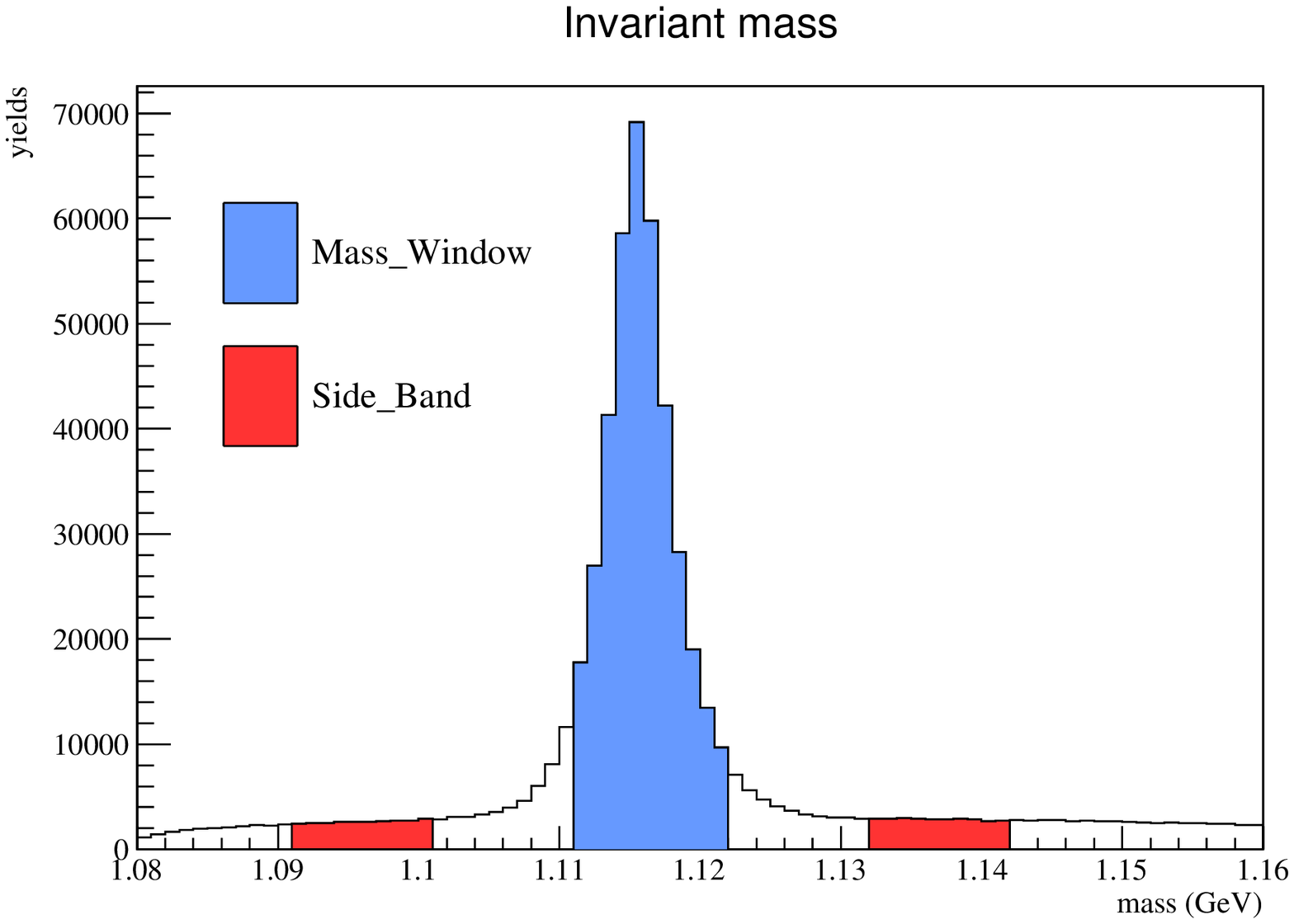}}
\caption{Invariant mass distributions of $\Lambda$ and $\bar{\Lambda}$ at $2<p_T<3$ GeV$/c$. 
The hyperon yields under the mass peak (blue filled area) are used for the $D_{LL}^{raw}$ calculation 
and the yields under the side-band region (red filled area) are used for estimating the background fraction under the hyperon mass peak.}
\label{fig:im_LA}
\end{figure}

\section{Hyperon Reconstruction and $D_{LL}$ Extraction}
In this measurement, the $\Lambda(\bar{\Lambda})$ hyperons are reconstructed via the weak decay channel $\Lambda \rightarrow p + \pi^{-}$ $(\bar{\Lambda}\rightarrow \bar{p} + \pi^+)$. The proton and pion candidate tracks measured by the Time Projection Chamber (TPC)\cite{TPC} are paired first, and then a set of topological selection criteria are applied to reduce the background fraction to the level about 10\% under the hyperon mass peak, as presented in \figurename{} \ref{fig:im_LA}.
To reconstruct the hyperon from the fragments of the outgoing parton, jets are first reconstructed using anti-$k_T$ algorithm\cite{anti-kT} with $R=0.6$ using the tracks measured by the TPC and the energy deposits in the Barrel and Endcap Electromagnetic Calorimeter (BEMC/EEMC)\cite{BEMC,EEMC}.
Then each hyperon is associated with a reconstructed jet by requiring $\Delta R = \sqrt{(\eta_{\Lambda(\bar{\Lambda})} - \eta_{jet})^2 + (\phi_{\Lambda(\bar{\Lambda})} - \phi_{jet})^2} < 0.6$. 

The extraction of $D_{LL}$ follows the same method as in \cite{DLL_2009,DLL_2018}.
The polarization of the $\Lambda(\bar{\Lambda})$ 
can be obtained via the angular distribution of the $\Lambda(\bar{\Lambda})$ decay products in the hyperon rest frame:

\begin{equation}
    \frac{dN}{d\mathrm{cos}\theta^*}=\frac{\sigma \mathcal{L}\text{ }\!A}{2}(1 + \alpha_{\Lambda(\bar{\Lambda})}P_{\Lambda(\bar{\Lambda})}\mathrm{cos}\theta^*),
    \label{eq:weak_decay}
\end{equation}
where the $\theta^*$ is the angle between the $\Lambda(\bar{\Lambda})$ momentum direction, i.e. longitudinal polarization direction, 
and the momentum of its daughter $p(\bar{p})$ at the $\Lambda(\bar{\Lambda})$ rest frame, and $A$ is the detector acceptance. 
The $\alpha_{\Lambda(\bar{\Lambda})}=0.732(-0.732)$\cite{PDG_2020} is the decay parameter 
and the $P_{\Lambda(\bar{\Lambda})}$ is the polarization of hyperons. 
To cancel the effect of the detector acceptance, the $D_{LL}$ is measured in small $\mathrm{cos}\theta^*$ intervals using the following equation \cite{DLL_2009}: 
\begin{equation}
    D_{LL}=\frac{1}{\alpha_{\Lambda(\bar{\Lambda})}P_{beam}\langle \cos\theta^* \rangle}\frac{N^+-\mathcal{R}N^-}{N^++\mathcal{R}N^-}.
    \label{eq:DLL_extract}
\end{equation}
Here, $N^+$($N^-$) denotes the hyperon yields with positive(negative) beam helicity, $P_{beam}$ is the beam polarization and $\mathcal{R}$ is the relative luminosity measured by the STAR Vertex Position Detector\cite{VPD}.

The raw spin transfers are extracted using the yields under the hyperon mass peak (blue area in \figurename{} \ref{fig:im_LA}) and are averaged over the whole $\mathrm{cos}\theta^*$ range. The residual background under the hyperon mass peak is mainly from random combinations of pion and proton candidates. This residual background is estimated using the side-band method (red area in \figurename{} \ref{fig:im_LA}). Its contribution to the $D_{LL}$ is corrected using the following equation:
\begin{equation}
    D_{LL}=\frac{D_{LL}^{raw}-rD_{LL}^{bkg}}{1 - r},
    \label{eq:DLL_raw-bkg}
\end{equation}
where $r$ is the residual background fraction under the hyperon mass peak.

\section{$D_{LL}$ Results}
In this section, we present the new preliminary results of $D_{LL}$ of $\Lambda$ and $\bar{\Lambda}$. Section 3.1 presents the results of $D_{LL}$ as a function of hyperon $p_{T}$, and Sec. 3.2 shows the first measurements on the $D_{LL}$ versus hyperon $z$ within jet. 

\subsection{$D_{LL}$ vs hyperon $p_T$}

\begin{figure}
    \centering
    \includegraphics[width=0.6\linewidth]{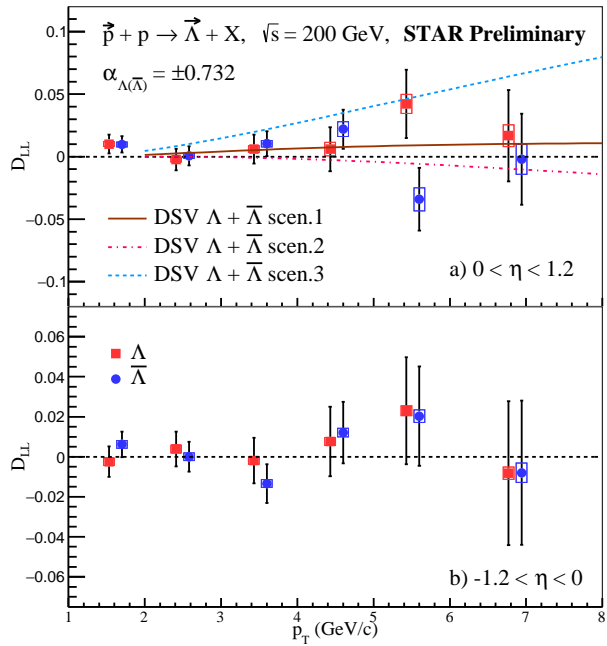}
    \caption{Preliminary results of $D_{LL}$ as a function of hyperon $p_T$ for $\Lambda$ and $\bar{\Lambda}$ in p+p collisions at $\sqrt{s}=200$ GeV. The top panel is for positive hyperon $\eta$ range and bottom one for negative $\eta$ range with respect to the polarized proton beam. The theoretical curves are from Ref. \cite{theory_work}.}
    \label{fig:DLL_vs_pT}
\end{figure}

Figure \ref{fig:DLL_vs_pT} shows the results of $D_{LL}$ as a function of hyperon $p_T$ in two different pseudo-rapidity ranges with respect to the momentum of the polarized proton. The results show consistency between $\Lambda$ and $\bar{\Lambda}$. The top panel shows the results for $0<\eta<1.2$. In this panel, the theoretical predictions\cite{theory_work} with different scenarios of the polarized fragmentation functions are compared with the measurements. The measurements are consistent with the model calculations within uncertainties.

\subsection{$D_{LL}$ vs $z$}
In this measurement, $z$ is defined as $z$ $ = \vec{p}_{\Lambda}\cdot \vec{p}_{jet}/|\vec{p}_{jet}|^2$, i.e. the longitudinal momentum fraction of the jet carried by the hyperon. The $z$ value at detector level can be obtained with the reconstructed jet, i.e. detector jet. Since the theoretical calculations consider all produced particles in a jet, then the detector-level $z$ needs to be corrected to the particle-level $z$. The correction is obtained with the Monte Carlo data based on the PYTHIA\cite{PYTHIA6.4} and Geant\cite{Geant3} simulation.

\begin{figure}
\centering
\includegraphics[width=0.6\linewidth]{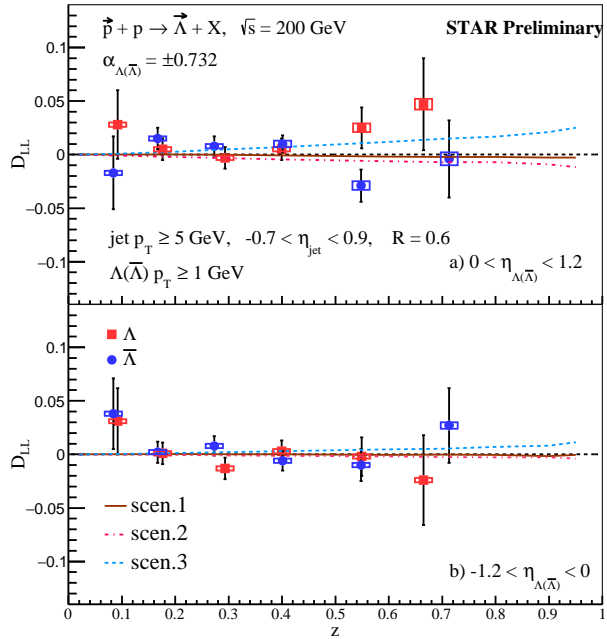}
\caption{Preliminary results of $D_{LL}$ as a function of hyperon $z$ for hyperon $p_T \geq 1.0$ GeV$/c$ and jet $p_T \geq 5.0$ GeV$/c$. 
The top panel is for positive hyperon $\eta$ range and the bottom one for negative $\eta$ range with respect to the polarized proton beam. The theoretical curves are from Ref. \cite{theory_work3} for three scenarios of polarized fragmentation functions.}
\label{fig:DLL_vs_z}
\end{figure}

The first results of $D_{LL}$ versus $z$ for $\Lambda$ and $\bar{\Lambda}$ in p+p collisions at $\sqrt{s} = 200$ GeV are presented in \figurename{} \ref{fig:DLL_vs_z}. The jet $p_T$ is required to be larger than 5 GeV$/c$. The top panel shows the results for the hyperon rapidity in the range of $0 < \eta < 1.2$ with respect to the momentum of the polarized beam and the bottom panel shows the results for $-1.2 < \eta < 0$. The results are compared with the theoretical predictions\cite{theory_work3} for three scenarios of polarized fragmentation functions\cite{theory_work6}. From \figurename{} \ref{fig:DLL_vs_z}, we can see that the results are consistent between $\Lambda$ and $\bar{\Lambda}$. The data are also consistent with the model calculations. However, current measurement precision does not yet allow for a clear discrimination of different scenarios.

\section{Summary}
In summary, new preliminary results on $\Lambda/\bar{\Lambda}$ longitudinal spin transfer $D_{LL}$ vs $p_T$, and $D_{LL}$ vs hyperon $z$ within a jet are presented for 200 GeV longitudinally polarized p+p collisions at STAR. 
These measurements provide insights into the polarized fragmentation functions for the $\Lambda$ and $\bar{\Lambda}$ hyperons 
and also the strange quark and anti-quark helicity distributions in the proton. 
The first measurement of the $D_{LL}$ vs $z$  in polarized p+p collisions directly probes the polarized fragmentation functions.

\section*{Acknowledgements}
The author is supported partially by the National Natural Science Foundation of China under No. 12075140.

\end{document}